\documentclass[preprint,prb,showpacs,preprintnumbers,amsmath,amssymb]{revtex4}
\usepackage{amssymb}

\usepackage{graphicx}
\usepackage{dcolumn}
\usepackage{bm}
\usepackage{epsfig}
\newcommand{\ignore}[1]{}

\begin{document}

\title{Formation of Hydrogenated Graphene Nanoripples by Strain Engineering and Directed Surface Self-assembly}

\author{Z. F. Wang}
\affiliation{Department of Materials Science and Engineering,
University of Utah, Salt Lake City, UT 84112, USA}

\author{Yu Zhang}
\affiliation{Department of Materials Science and Engineering,
University of Utah, Salt Lake City, UT 84112, USA}

\author{Feng Liu}\thanks{Corresponding author. E-mail: fliu@eng.utah.edu}
\affiliation{Department of Materials Science and Engineering,
University of Utah, Salt Lake City, UT 84112, USA}

\begin{abstract}
We propose a new class of semiconducting graphene-based
nanostructures: hydrogenated graphene nanoripples (HGNRs), based on
continuum-mechanics analysis and first-principles calculations. They
are formed via a two-step combinatorial approach: first by strain
engineered pattern formation of graphene nanoripples, followed by a
curvature-directed self-assembly of H adsorption. It offers a high
level of control of the structure and morphology of the HGNRs, and
hence their band gaps which share common features with graphene
nanoribbons. A cycle of H adsorption/desorption at/from the same
surface locations completes a reversible metal-semiconductor-metal
transition with the same band gap.
\end{abstract}


\maketitle

Nanostructures have distinct properties from their bulk
counterparts. In the case of graphene, nanostructuring affords an
effective mean to convert the semimetal graphene \cite{1} into
semiconducting graphene based nanostructures, which is desirable for
many nanoelectronics applications. A number of theoretical proposals
and experimental attempts have been made to create graphene based
nanostructures, such as graphene nanoribbons \cite{5,6,7,13,13a},
nanohole superlattices \cite{8,9,14}, hydrogenated graphene
nanostripes \cite{10,11,15} and graphane \cite{graphane}. However,
our current success is still far below our expectations. Although the
physical principles for opening graphene band gap are well established,
the synthesis of the semiconducting graphene-based nanostructures with desirable
precision and control remains challenging. Lithographic patterning
of graphene into nanodimensions has difficulties in controlling the
nanopattern size and edge qualities. The method using H adsorption
on graphene is fundamentally a stochastic process, and how to direct
H to the exact locations as needed is not established.

In this Letter, we propose a strain engineered self-assembly process
to form a new class of graphene-based nanostructures, the
hydrogenated graphene nanoripples (HGNRs). The process consists of
two steps: first strain engineering graphene into periodic
nanoripple patterns, followed by a directed self-assembly of H
adsorption onto the nanoripple template. The combination of the
strain engineering and the directed H surface self-assembly offers a
high level of control of the dimensions of the HGNRs, and hence
their band gaps which share the common scaling features with
graphene nanoribbons.

Generally, two physical mechanisms have been employed for opening
band gap in graphene. One is by imposing the quantum confinement
effect. The semimetal behavior of graphene stems from the free
motion of 2D $\pi$ electrons. If the motion of $\pi$ electrons is
confined, then band gap opens. This can be achieved by cutting
graphene into nanoribbons \cite{5,6,7,13} and nano-networks
\cite{8,9,14}, or by H adsorption \cite{10,11,15} where locally
$\pi$ bands are removed due to the change from $sp^2$ to $sp^3$
hybridization. The other mechanism is by breaking the graphene
lattice symmetry. In the pristine graphene, the $\pi$ bands,
residing on A-sub lattice, are degenerate with the $\pi^*$ bands,
residing on B-sub lattice, at Fermi energy. If such symmetry is
broken \cite{17}, then gap opens. This has been shown in the
spin-polarized graphene zigzag edges \cite{6,7,10,11} and in the
epitaxial graphene grown on compound BN \cite{12} and SiC \cite{16}
substrates, where spin and substrate potential lifts the energy
degeneracy between the $\pi$ and $\pi^*$ bands, respectively.

\begin{figure}[htpb]
\begin{center}
\epsfig{figure=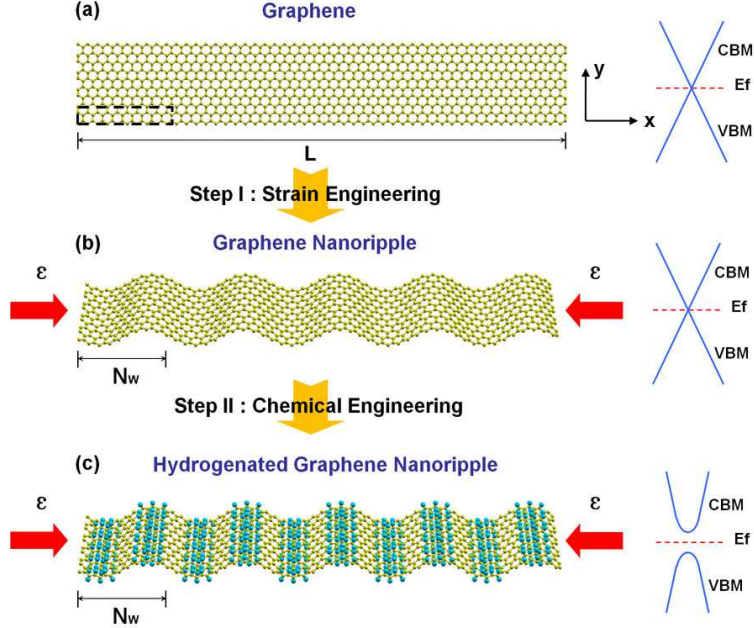,width=10cm} \caption{Schematic illustration
of a combinatorial approach of straining graphene and directing H
adsorption to form HGNR. (a) Pristine graphene with zero band gap.
$L$ denotes the length. The dashed line shows the unit cell.
(b) Graphene nanoripple with zero band gap.
$\varepsilon$ and $N_W$ denote the compressive strain and the ripple
period, respectively. (c) HGNR with non-zero band gap. $E_f$ is the
Fermi energy; CBM (VBM) is the conduction (valence) band minimum
(maximum).}
\end{center}
\vspace{-0.2in} \label{fig:fig-1}
\end{figure}

To implement the above mechanisms, the basic idea of our
combinatorial approach of strain engineering and directed surface
self-assembly is illustrated in Fig. 1. Starting with a pristine
graphene sheet of length $L$ [Fig. 1(a)], in the first step of
"strain engineering", a compressive strain ($\varepsilon$) is
applied along $L$ to form a 1D pattern of graphene nanoripple with
period $N_W$ [Fig. 1(b)] [$N_W$ is the number of dimer and
zigzag lines denoting the width of armchair and zigzag nanoripples,
respectively (see Fig. 3(a) and 3(d))]. At this step, the
nanoripple remains a semimetal. In the second step of "chemical
engineering", H atoms are introduced to adsorb onto the ripple
pattern at designated locations [Fig. 1(c)] to form the HGNRs. Now,
band gap opens. The nanoripple formation in the first step has two
key functions. First, its period defines the period of the HGNR,
\emph{i.e.} the width of the hydrogenated nanostripes created in the
second step, and hence the eventual size of band gap. Second, its
morphology serves as a template to direct H atoms to be adsorbed at
specific locations with the largest local curvature, so as to form
highly ordered H patterns. The directed H adsorption in the second
step fulfills the role to open a predefined band gap.

There are several noteworthy advantages of the above approach. The
band gap as engineered is tunable with a high level of control. It
is well-known that the band gap scales inversely with graphene
nanoribbon widths \cite{5,6,7,13}. Here, the width of HGNR is
uniquely defined by the period of the nanoripple without H
adsorption, which can be tuned precisely by the magnitude of the
compressive strain applied and the length of graphene used. Because
the H atoms are directed by the nanoripple template to the
designated locations of the largest curvature, instead of random
adsorption sites, they form a highly regular pattern, which
translates the HGNR into an ordered array of graphene nanostripes
with the highly uniform width, orientation and smooth edges, so that
they all open a predefined uniform band gap. Furthermore, the
approach makes a repeatable process, as a cycle of the directed H
adsorption and desorption leads naturally to a cycle of
metal-semiconductor-metal transition opening the same band gap in
the HGNR. In the following, we discuss the processing paramters of
the HGNRs and their resulting electronic properties.

We first analyze the strain induced nanoripple pattern formation in
graphene, based on a continuum mechanics model \cite{18,zhang}. Consider a
uniaxial compressive strain $\varepsilon$ applied along the
x-direction [Fig. 1(b)] of a graphene sheet of length $L$. Above a
critical strain value, the flat graphene becomes unstable and
undulates into a 1D sinusoidal ripple pattern with period $N_W$. The
total energy of the ripple can be calculated as
\begin{eqnarray}
U_{total}&=&U_{bending}+U_{streching}\\
\notag
&=&\frac{B}{2}\int(\frac{\partial^2\xi}{\partial{x^2}})^2dxdy+
\frac{Eh}{2}\int\varepsilon(\frac{\partial\xi}{\partial x})^2dxdy
\end{eqnarray}
where $B$ and $E$ are the bending and Young's modulus of graphene,
respectively, $\xi$ is the displacement along the z direction, and
$h$ is the thickness of graphene. Under the boundary condition
$\frac{\partial^2\xi}{\partial{x^2}}|_{x=0,x=L}=0$,
$\xi(x)=\frac{C}{(n\pi/L)^2}sin(\frac{n\pi x}{L})$, where $n$ is the
number of ripple period formed and $C$ is a constant. Applying the
variational principle, the critical strain for generating the ripple
pattern can be obtained as
\begin{eqnarray}
\varepsilon_{cr}=\frac{Bn^2\pi^2}{EhL^2}=\frac{h^2n^2\pi^2}{12(1-\nu^2)L^2}
\end{eqnarray}
where $\nu$ is the Poisson's ratio.

\begin{figure}[htpb]
\begin{center}
\epsfig{figure=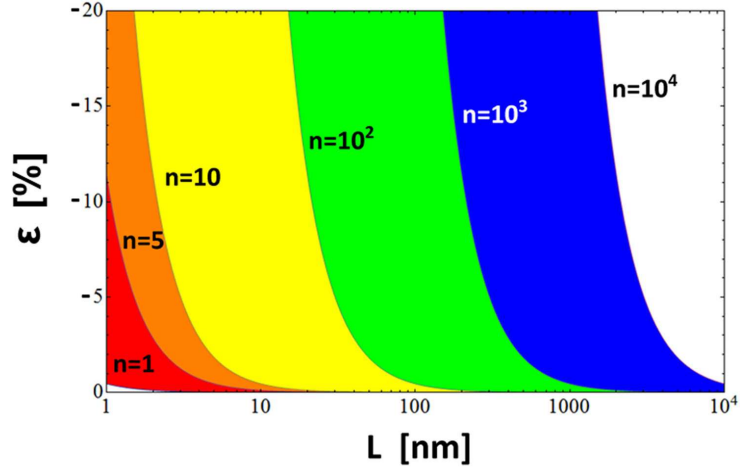,width=10cm} \caption{Phase diagram showing the number of
ripple period ($n$) formed as a function of the graphene length ($L$) and the applied
compressive strain ($\varepsilon$). The boundary lines mark the critical strain
($\varepsilon_{cr}$) in Eq. (2).}
\end{center} \vspace{-0.2in} \label{fig:fig-2}
\end{figure}

\begin{figure}[htpb]
\begin{center}
\epsfig{figure=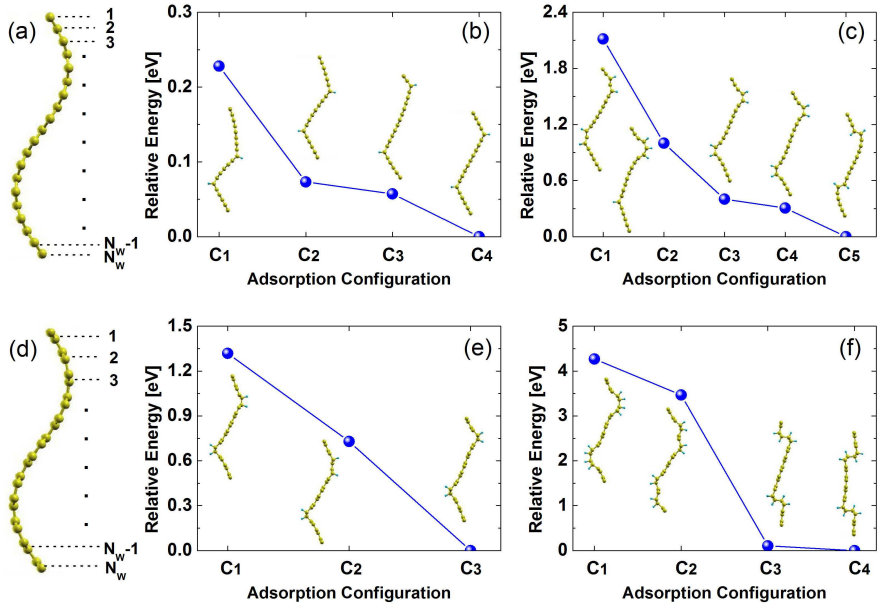,width=12cm} \caption{Directed H adsorption
onto the graphene nanoripple. (a) and (d)  Schematics of one period
of armchair and zigzag nanoripple, respectively. $N_W$ denotes the
ripple period. (b) and (c) Relative energies of different H
adsorption configurations ($C_1$ through $C_5$, shown as the
inserts) of armchair HGNR with two and four rows of H atoms
within one period of the ripple, respectively. (e) and (f) Same as
(b) and (c) for zigzag HGNR. Note that the H row in armchair HGNR is
a straight line and in zigzag HGNR a zigzag line.}
\end{center} \vspace{-0.2in} \label{fig:fig-3}
\end{figure}

Figure 2 shows the calculated phase diagram showing the number of ripple
period ($n$) as a function of strain ($\varepsilon$) and graphene length ($L$),
using $\nu=0.34$ \cite{19} and $h=0.7{\AA}$ \cite{20}. For the given $L$ ($\varepsilon$),
$n$ increases with the increasing $\varepsilon$($L$). Thus, the period of ripple
pattern ($N_W$) can be tuned by the magnitude of the compressive strain and the
length of graphene.

Next, we analyze a "directed" H self-assembly on the ripple pattern.
The ripple morphology provides a curvature template to direct the H
adsorption to the designated locations. To illustrate this effect,
we compare the total energies of different hydrogen adsorption
configurations using the first-principles method. The calculations
are performed using the VASP package \cite{21}, which implements the
local (spin) density approximation \cite{22} of the density
functional theory. The electron-ion interaction is described by the
projected augmented wave method \cite{23} with an energy cutoff of
$400 eV$ and a k-point mesh of ($15\times7\times1$) based on the
convergence tests. The atom positions are optimized with the atomic
force converged to less than $0.01 eV/{\AA}$.

There are two typical orientations in graphene, \emph{i.e.} the
zigzag and armchair direction \cite{5,6}, along which we apply the
compressive strain to form the corresponding armchair and zigzag
nanoripple, as shown in Figs. 3(a) and (d), respectively. For the
armchair nanoripple, we choose $N_W=20$ and $\varepsilon=-10\%$ as an
example to demonstrate the directed H adsorption. First, we adsorb
two rows of H atoms within one period of the nanoripple to form the
armchair HGNR. Comparing the relative adsorption energies
(setting the energy per unit cell of the most stable
adsorption configuration as the zero energy of reference)
[Fig.3(b)], we can see clearly that the H atoms prefer to adsorb
onto the carbon atoms with the largest curvature. The H rows divide
one ripple period into two equivalent nanostripes of the same width.
Next, we absorb another two rows of H atoms, and the stable H
adsorption configuration is shown in Fig. 3(c). The additional H
rows are attracted to the existing rows, forming two rows of H at
the largest curvature locations. Similar directed H adsorption are
also observed in the zigzag HGNR with $N_W=12$ and
$\varepsilon=-10\%$, as shown in Figs. 3(e) and (f) for two rows and
four rows of H atoms within one period of zigzag nanoripple,
respectively.

The preferred H adsorption onto the C atoms with the largest curvature is
consistent with the previous theoretical \cite{24} and experimental \cite{25} results.
This is because instead of the $sp^2$ electronic configuration in a planar symmetry,
the curved C atom has a $sp^{2+\delta}$ configuration, which is closer to the final $sp^3$
configuration upon H adsorption \cite{26}. Thus, it costs less energy for H to adsorb
onto a curved C atom than onto a planar C atom. Based on the same principle, after H
is adsorbed on a C atom, it makes its neighboring C atoms more $sp^{2+\delta}$ like, so
that the additional H atoms will prefer to adsorb onto these neighboring C atoms, but
on the opposite side of graphene.

\begin{figure}[htpb]
\begin{center}
\epsfig{figure=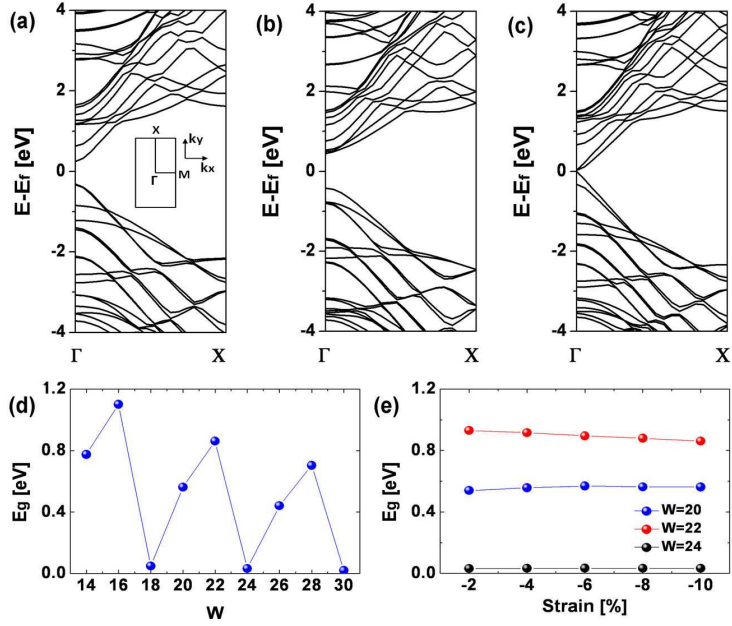,width=10cm} \caption{Band Structure of the
armchair HGNR with two rows of H atoms. (a)-(c) Band structures for
$N_W=20$, $22$ and $24$ with $\varepsilon=-10\%$, respectively.
The inset in (a) shows the first Brillouin zone with three
high-symmetry points and the reciprocal coordinate axes.
(d) Band gap as a function of $N_W$
for the fixed strain $\varepsilon=-10\%$. (e) Band gap as a function
of strain for $N_W=20$, $22$ and $24$.}
\end{center} \vspace{-0.2in} \label{fig:fig-4}
\end{figure}

\begin{figure}[htpb]
\begin{center}
\epsfig{figure=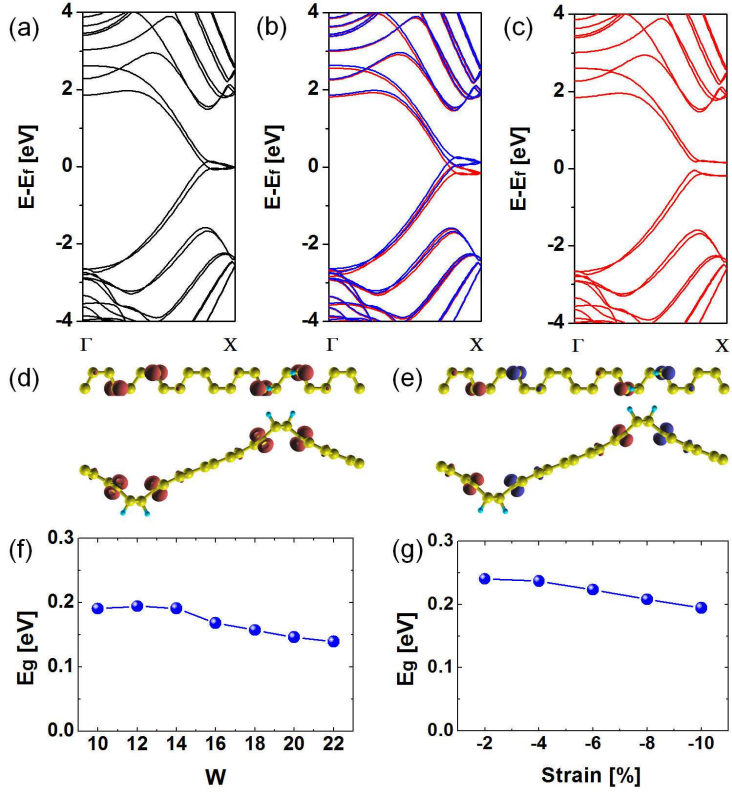,width=10cm} \caption{Band Structure of the
zigzag HGNR with two rows of H atoms. (a)-(c) Band structures for
the NM, FM and AFM states with $N_W=12$ and $\varepsilon=-10\%$,
respectively. The high-symmetry points have the same meaning
as those in Fig. 4(a). The red (blue) color corresponds to spin-up
(spin-down). Spin-up and spin-down is degenerate in (c). (d) and (e)
The spatial distribution of spin density
($\rho_{spin-up}-\rho_{spin-down}$) for FM (isovalue 0.04) and AFM
states (isovalue 0.02) with $N_W=12$ and $\varepsilon= -10\%$,
respectively. The color has the same spin notion as in (b). (f)
Band gap of the AFM state as a function of $N_W$ for the fixed strain
$\varepsilon= -10\%$. (g) Band gap of the AFM state as a function of
strain for $N_W=12$.}
\end{center} \vspace{-0.2in} \label{fig:fig-5}
\end{figure}

Formation of graphene nanoripple without H doesn't open band gap;
one role of ripple structure is to provide a template to direct the
H adsorption that will open band gap. Also, the period of the
nanoripple patterns defines the period of the HGNR, \emph{i.e.} the
width of nanostripes formed upon H adsorption, and hence the final
band gap. Next, we present the band gaps of the HGNRs as a function
of nanoripple period and strain.

In the direction perpendicular to the H row ($\Gamma M$
direction), the band structures for armchair and zigzag HGNRs are
almost flat, so in the following, we only show the band structures
along the H row direction ($\Gamma X$ direction). The band
structures of armchair HGNR adsorbed  with two rows of H atoms are
shown in Fig. 4, similar to those of armchair graphene nanoribbons
\cite{5,6}. Figures 4(a)-(c) are the respective band structures for
$N_W=20$, $22$ and $24$ with $\varepsilon=-10\%$, in which band gaps
can be clearly seen. Without the H, no band gap appears in the
armchair nanoripples even when the strain goes up to $-30\%$.
The physical origin of the band gap is due to quantum
confinement, \emph{i.e.} the adsorbed H rows remove local $\pi$
bands confining electrons between them. For the fixed strain, the
band gap is not a monotonous function of the ripple width, and can
be divided into three groups (Fig. 4(d)). They follow the relation,
$E_g(N_{W_{eff}}=3p+2)<E_g(N_{W_{eff}}=3p)<E_g(N_{W_{eff}}=3p+1)$, which is same
as that for the armchair graphene nanoribbons \cite{6}. Here
$N_{W_{eff}}=(N_W-2)/2$ is the effective nanostripe width and $p$ is a
positive integer. For the fixed ripple width, the three band gap
groups show little dependence on the strain [Fig. 4(e)]. Without
ripple formation, the band gaps of flat armchair nanoribbons have
been shown to depend on in-plane uniaxial strain \cite{27}, because
strain changes bond length and hence the interatomic electron
hopping energies. With the ripple formation, however, the in-plane
strain is largely relaxed by the bending (\emph{i.e.} change of bond
angles) so that the bond length changes very little. This is the
reason we see a very weak strain dependence of band gap in Fig.
4(e). Note that under compressive strain, the ripple structure is
usually favored over the planar graphene, because for normal
graphene size ($L \sim 10-10^4nm$), the critical strain for ripple
formation is extremely small ($<0.1\%$), as shown in Fig. 2.

The band structures of zigzag HGNRs adsorbed with two rows of H
atoms are shown in Fig. 5, similar to those of zigzag graphene
nanoribbons \cite{5,6}. Figures 5(a)-(c) show the band structures of
the nonmagnetic (NM), ferromagnetic (FM) and antiferromagnetic (AFM)
states with $N_W=12$ and $\varepsilon=-10\%$, respectively. Same as
zigzag nanoribbons, the zigzag HGNR in Fig. 5 has an AFM ground
state, with the NM and FM states being $25.9meV$ and $13.8meV$
higher in energy than the AFM state. The NM state is a metal, with
four subbands crossing the Fermi level [Fig. 5(a)]. This is
different from the zigzag nanoribbon, which has only two subbands
crossing the Fermi level \cite{6}. Because the H atoms divide one
ripple period into two nanostripes of the same width, their
interaction splits the two subbands into four. The FM state is a
metal, while the AFM state is a semiconductor for both spins. The
spatial distribution of spin charge densities are shown in Figs.
5(d) and (e) for the FM and AFM states, respectively. For the fixed
strain, the band gap of the AFM zigzag HGNR [Fig. 5(f)], opened by
symmetry breaking mechanism, shows a much weaker dependence on the
ripple width compared to the band gaps of armchair HGNR [Fig. 4(d)],
opened by quantum confinement mechanism. For the fixed ripple width,
the band gap of the AFM zigzag HGNR [Fig. 5(g)] shows a rather weak
dependence on the strain, similar to the case of armchair HGNR [Fig.
4(e)].

In conclusion, we demonstrate theoretically a strain engineered
self-assembly process to fabricate a new class of semiconducting
graphene based nanostructures, the HGNRs, which share some common
band-gap features of graphene nanoribbons. It is a combinatorial
two-step process of straining graphene sheet into nanoripples
followed by the curvature-directed H adsorption, which offers a high
level of band-gap control by tuning the magnitude of strain, the
dimension of graphene sheet and the amount of H adsorption. We note
that graphene nanoripples have already been fabricated by straining
suspended graphene sheet experimentally \cite{28}. The prospect of
further dosing  the nanoripples with H to form the semiconducting
HGNRs is very appealing. We expect that the combination of strain
engineered nanoripple formation and the curvature-directed surface
self-assembly can be generally applied beyond graphene to other
nanomembranes.

This work was supported by DOE, CMSN and BES programs. We thank
Center for High Performance Computing at the University of Utah for
providing the computing resources.

\end{document}